\documentclass[11pt]{article}

\usepackage[final]{acl}

\usepackage{times}
\usepackage{latexsym}
\usepackage{amsmath}
\usepackage{enumitem}

\usepackage{arydshln}
\usepackage{xcolor}

\usepackage[T1]{fontenc}

\usepackage[utf8]{inputenc}

\usepackage{microtype}

\usepackage{inconsolata}

\usepackage{graphicx}

\usepackage{mdframed}
\usepackage{multirow}
\usepackage{multicol}

\title{Bridging Business Intent and Data: A Benchmark for Automatic Relational Data Product Generation}

\author{
  \textbf{Faisal Chowdhury$^*$},
  \textbf{Sola Shirai$^*$},
  \textbf{Sarthak Dash$^*$}, \\
  \textbf{Nandana Mihindukulasooriya},
  \textbf{Horst Samulowitz}\\
  IBM Research, Yorktown Heights, USA\\
  \texttt{\{mchowdh, sdash, samulowitz\}@us.ibm.com}, \texttt{\{solashirai, nandana\}@ibm.com}
}

\begin{document}
\maketitle


\begin{abstract}

A data product is designed to address a specific business need by transforming raw data into a curated, usable asset that delivers actionable insights. Despite practical advances in related areas like text-to-SQL and ELT pipelines, there is no comprehensive benchmark for evaluating the end-to-end process of automatically generating such data products from high-level business requests.
To fill this gap, we introduce DP-Bench, a first-of-its-kind benchmark for automatic data product creation, built by combining elements from existing ELT and text-to-SQL datasets. We also propose baseline methods using LLMs to generate data products, providing a foundation for future research in bridging natural language business intent and structured data representation. We make the DP-Bench dataset and code available at: \url{https://huggingface.co/datasets/ibm-research/dp-bench}.

\end{abstract}

\section{Introduction}

The concept of a \textit{data product} was first introduced by \citet{patil2012} as ``a product that facilitates an end goal through the use of data.'' It represents a curated and structured data asset designed to address a specific analytical objective or business use case, such as predicting customer churn or analyzing sales trends. Data products have gained renewed attention as a foundational component of the data mesh paradigm~\cite{patil2012}, where domain-specific teams are responsible for building discoverable, trustworthy, and tailored data assets.

Despite significant advances in related areas—such as text-to-SQL, schema linking, and ELT pipeline generation—existing approaches focus on isolated components of the data lifecycle. Real-world data product creation is inherently holistic: it begins with a high-level business request and culminates in a structured dataset that includes relevant schema elements, derived features, and explicit transformation logic. Yet no standardized benchmark exists for evaluating systems that perform this end-to-end transformation.

To address this gap, we introduce \textbf{DP-Bench} (Data Product Benchmark). It consists of natural language \textit{Data Product Requests} (DPRs), corresponding data products (structured schema subsets with base and derived columns), and transformation provenance expressed in SQL. Every component is manually curated and validated by human annotators. We also propose baseline approaches for automatic data product generation. We refer to our Limitations section for scope boundaries.

\section{Formal Task Definition}

We formalize the task of automatic data product creation as a structured prediction problem over database schemas given a natural language data product request (DPR).

\subsection{Data Product Request — DPR}
\label{dpr_def}
Formally, let $r \in \mathcal{R}$ denote a \emph{Data Product Request (DPR)}: a natural language description capturing a business problem, analytical goals, and implicit requirements for relevant data entities and transformations. A DPR is \textbf{goal-oriented} (specifying \emph{what} rather than \emph{how}), \textbf{abstractive} (unifying a set of analytical queries into a business intent), and \textbf{schema-groundable} (mappable to elements of an underlying schema).

\subsection{Data Product}
Let $\{DB\}$ denote a relational database associated with a schema $\mathcal{S}$ consisting of a set of tables, columns, primary keys, and foreign key relations.

A \emph{data product} $DP$ corresponding to a request $r$ is defined as a tuple:
\[
DP = (T, C, C_d, P)
\]
where $T$ is a set of selected tables from the schema, $C$ is a set of selected (non-derived) columns from the tables in $T$, $C_d$ is a set of derived columns constructed from base columns in $C$, and $P$ is a set of provenance functions. Each $p \in P$ is uniquely associated with one derived column $c_d \in C_d$ and defines the transformation used to compute it from the base columns (e.g., expressed as a SQL query).


\subsection{Problem Formulation}
Given $r$ and database $DB$ (schema $\mathcal{S}$), the goal is to learn $f: (r, \mathcal{S}) \rightarrow \widehat{DP}$ that maximizes $\mathcal{L}(\widehat{DP}, DP^*)$, a composite of schema selection accuracy, derived column correctness, and downstream task performance. This formulation generalizes schema linking and text-to-SQL into a more expressive task.

\subsection{Constraints}
Generated DPs must satisfy: \textbf{Schema Validity} (all selected tables/columns exist in $\mathcal{S}$), \textbf{Transformation Validity} (derived columns are computable from available base columns), and \textbf{Provenance Availability} (each derived column has a valid transformation function).

\subsection{Example}
\begin{figure*}[ht]
  \centering
  \includegraphics[width=1.05\textwidth]{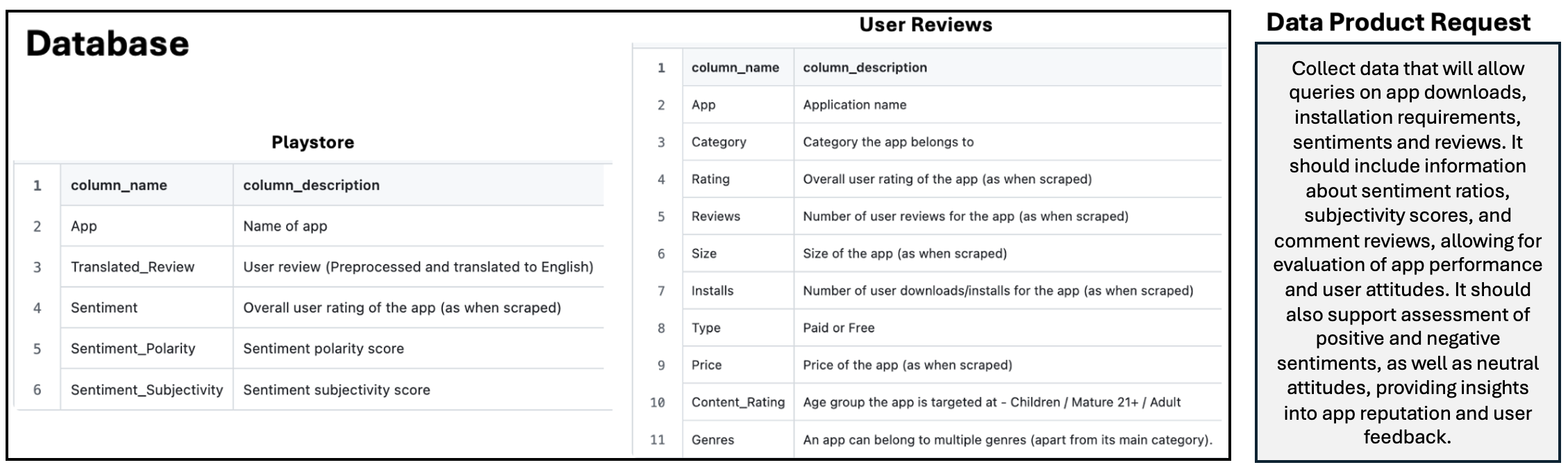}
  \caption{\centering Example of an input DB and a data product request for this DB.}
  \label{fig:example_dp_input}
\end{figure*}

Figure \ref{fig:example_dp_input} shows an input database with a DPR describing analytical needs over mobile applications.

\begin{figure}[h]
  \centering
  \includegraphics[width=0.5\textwidth]{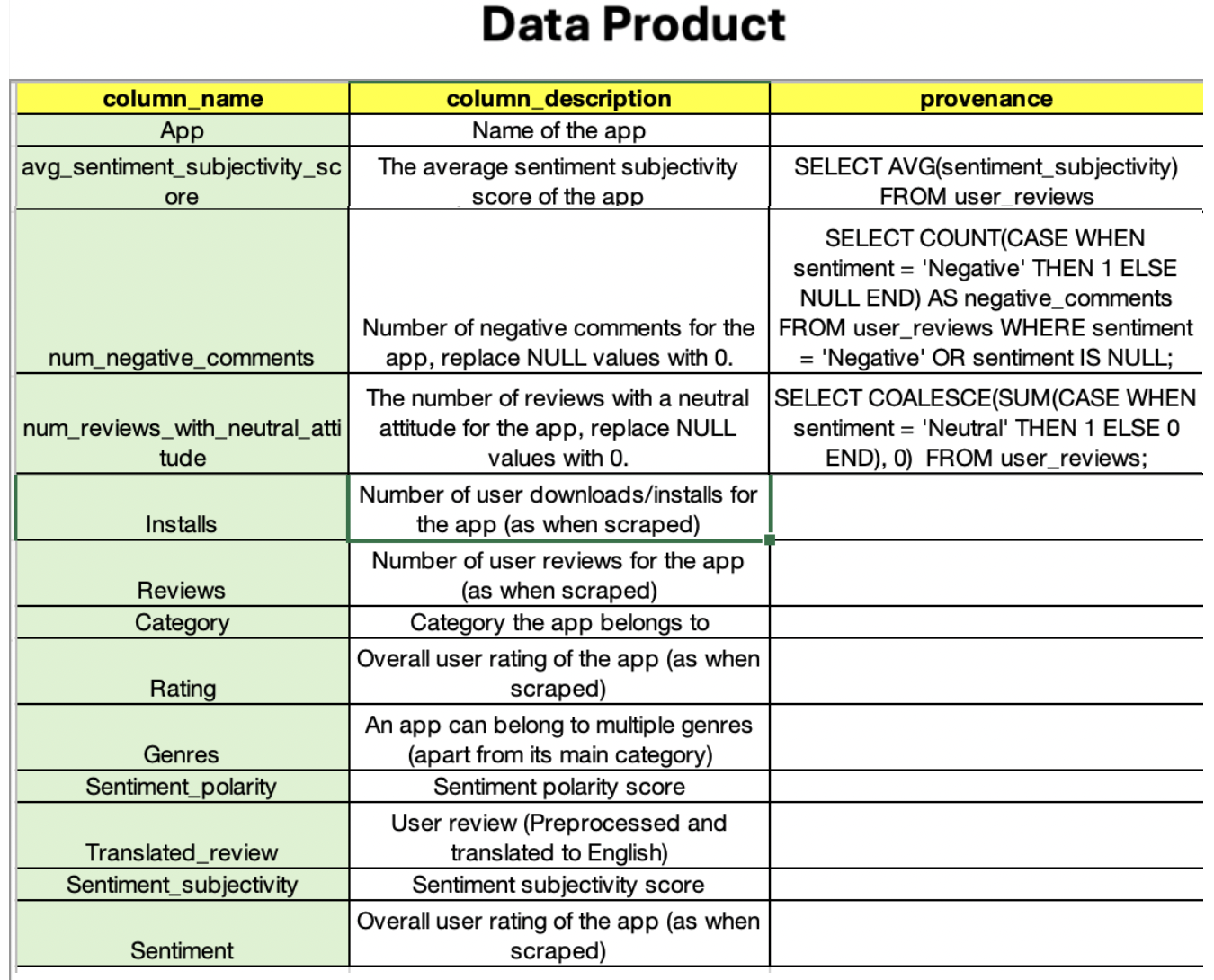}
  \caption{\centering Example of a DP for the input of Figure \ref{fig:example_dp_input}.}
  \label{fig:example_dp}
\end{figure}

Figure \ref{fig:example_dp} shows the corresponding data product. Base columns such as \texttt{App} and \texttt{Installs} are directly selected; derived columns (e.g.\ \textit{``num\_negative\_comments''}) are constructed via SQL transformations. Columns not derivable from the available schema are excluded.

\section{Data Sources for DP-Bench}

\textbf{BIRD}~\cite{bird} is a cross-domain English text-to-SQL benchmark covering 37+ domains with 12,751 question--SQL pairs for 95 databases. \textbf{ELT-Bench}~\cite{elt-bench} targets ELT pipeline construction and consists of 100 pipelines across 100 databases, with 835 source tables and 203 data models. Each data model contains English names, descriptions, and column metadata; columns may be mapped directly to a source table or derived from multiple tables. Examples are shown in Appendix \ref{apen-bird-elt-bench}.

\section{DP-Bench Construction}

To construct DP-Bench, we re-purposed ELT-Bench and BIRD using LLM-based approaches (Llama-3.3 70B \cite{grattafiori2024llama3herdmodels}) for column alignment and request generation, with all outputs manually vetted by the five human annotators. We first annotated the expected outputs (i.e.\ the data products) and then worked backwards to annotate the appropriate inputs (i.e.\ data product requests).

\subsection{Align BIRD and ELT-Bench} 


Out of 100 ELT-Bench pipelines, \textbf{78} can be mapped one-on-one to BIRD databases. These 78 BIRD databases together contain \textbf{4,306} columns overall and are associated with \textbf{10,928} unique question-SQL pairs. The 78 ELT-Bench pipelines have \textbf{921} columns within \textbf{122} data models.

We used an LLM to align columns within data models of ELT-Bench to the columns within tables of BIRD. This analysis discovered that \textbf{634} (out of 921) columns within ELT-Bench's data models are \textbf{derived} from columns within BIRD databases. Consequently, the remaining \textbf{287} (out of 921) columns are an \textit{exact} match to columns within BIRD. Henceforth, we shall refer to these columns as \textbf{non-derived} columns.

\subsection{Produce Provenance for Derived Columns}

We used the prompt in Appendix \ref{apen-provenance} to generate provenance in SQL for the derived columns.

We then manually validated whether the generated SQL provenance for a derived column can be used to populate desired values for that column.

During this step, we had to filter out some derived columns due to erroneous provenance which is described with example in Appendix \ref{app:der_col_fil}. The total number of remaining derived columns is \textbf{582}.

\subsection{Create Preliminary Data Products}
\label{subsec:create-dp}

We create one data product for each of the 78 databases by grouping the \textbf{582} derived columns and \textbf{287} non-derived columns by BIRD table name. For each ELT-Bench data model, we created a table in the data product with the aligned BIRD table name, automatically adding non-derived columns and manually reviewing derived column placement.

\subsection{Annotate Data Product Requests (DPRs)}
\subsubsection{\bf Candidate DPR Generation}
We coin the term Data Product Request (DPR) for targeted business-oriented natural language queries an \textit{analyst user persona} can ask against a given data product.

To build a DPR (Section \ref{dpr_def}), we use a two-stage strategy. First, given the tables of a data product, we use an LLM to generate 3 possible summaries using the prompt in Appendix \ref{apen-dp-summary}. Second, given each summary, we use the LLM to convert it into a \textbf{candidate data product request} (Appendix \ref{apen-dpr}). Three summaries per DP yield \textbf{234} candidate DPRs for 78 data products.

\subsubsection{\bf Manual Curation for Gold DPRs}
We performed a 2-phase curation of the 234 candidate DPRs. In phase 1, each DPR was assessed by 2 annotators who verified—using 5 randomly chosen BIRD questions per data product—whether the DPR adequately captures the business intent needed to answer those questions; if not, they edited it. In phase 2, a 3rd annotator adjudicated any conflicting edits. Overall, 71\% of LLM-generated DPRs required no edits; 67 DPRs across 27 data products were revised.

\subsection{Enhance Preliminary Data Products}
After SQL error checking and filtering, we found \textbf{9,218} (84\%) of the previously mentioned 10,928 unique questions--SQL pairs from the aforementioned 78 databases in BIRD are noise free. The column names and table names used in these SQLs can be traced to tables in corresponding databases. We used \emph{sql\_metadata}\footnote{\url{https://pypi.org/project/sql-metadata/}} for parsing SQL. For each DPR, we select questions with Granite embedding cosine similarity $\geq$0.8, yielding \textbf{836} relevant questions. Tables and columns from their SQLs are added to the corresponding DPs, giving a final benchmark of \textbf{1,458} non-derived and \textbf{582} derived columns in \textbf{383} tables across \textbf{78} data products (distribution in Appendix \ref{apen-dp-details}). DBs with $\geq$50 columns form the \textbf{Hard} subset (30 DBs).

\subsection{Annotate topics}
\label{subsec:anno_topics}
Topics provide a higher-level abstraction over DPs to enable indexing, semantic alignment evaluation, and DPR--DP comparison. We annotated \textbf{564} topics across 78 data products; details and experiments are in Appendix \ref{app:topic} and \ref{app:topic-exp}.

\section{Study Design and Evaluation Metrics}

We evaluate systems along three dimensions: (i) schema selection accuracy, (ii) derived column generation quality, and (iii) downstream text-to-SQL effectiveness.

\subsection{Research Questions}
Our evaluation is guided by the following four research questions:
\textbf{RQ1:} How accurately can systems select relevant tables and columns for a given DPR?
\textbf{RQ2:} How well can systems generate derived columns with correct transformations and provenance?
\textbf{RQ3:} How effectively do generated data products support downstream analytical tasks (e.g., text-to-SQL)?
\textbf{RQ4:} How does database complexity affect performance?

\subsection{Evaluation Overview}
Given a DPR, a ground-truth DP, and a predicted DP, we evaluate systems using the following metrics.

\subsubsection{Column-Level Metrics}
We measure the overlap between predicted and ground-truth schema elements using Precision (P), Recall (R), and F1 score over selected columns (and tables implicitly). Formally,
\[
P = \frac{|\hat{C} \cap C^*|}{|\hat{C}|}, \quad
R = \frac{|\hat{C} \cap C^*|}{|C^*|}, \quad
F1 = \frac{2PR}{P + R}
\]
where $\hat{C}$ and $C^*$ denote predicted and ground-truth column sets, respectively.

We consider two settings for these metrics. The first is the \textit{non-derived} setting, where we only consider $\hat{C}$ which are base (non-derived) columns. The second is the \textit{inclusive} setting, where $\hat{C}$ also includes the columns which are used to produce derived columns. The columns that make up these two settings can differ because a DP may use columns to produce derived columns that are relevant to the use-case but not select the base columns on their own.

\subsection{Derived Column Similarity}
To evaluate generated derived columns, we compute similarity between transformation definitions using abstract syntax trees (ASTs). Given two derived columns with ASTs $x$ and $y$, similarity is defined as:
\[
\text{sim}(x,y) = 1 - \frac{\text{diff}(x,y)}{\text{count}(x,y)},
\]
where $\text{diff}(x,y)$ is the edit distance between ASTs and $\text{count}(x,y)$ is the total count of comparisons made between the ASTs, which normalizes $\text{sim}(x,y)$ to range from 0 to 1. Derived columns from predicted and ground-truth DPs are then matched to maximize total similarity, and which we use to compute soft Precision, Recall, and F1.

\subsection{Downstream Task Evaluation}
We evaluate the usefulness of generated data products using a text-to-SQL task. For each DPR, we measure execution accuracy:
\[
\text{Execution Accuracy} = \frac{\# \text{correct query results}}{\# \text{queries}},
\]
where correctness is determined by comparing outputs of generated SQL queries with ground-truth.

Column-level metrics assess schema selection (RQ1), derived column similarity addresses transformation quality (RQ2), and execution accuracy captures end-to-end utility (RQ3). Evaluation across databases of varying sizes addresses RQ4.

\section{Baselines for Automatic Data Product Creation}

\subsection{Hybrid Search}

The hybrid search baseline builds a hybrid index of dense (sentence transformer~\cite{reimers-gurevych-2019-sentence}) and sparse (BM25~\cite{robertson2009probabilistic}) embeddings of table--column sentences. The DPR is embedded similarly; combined similarity scores (weights 0.8/0.2) rank columns. Columns above a threshold of 0.75 are selected; foreign-key columns are added when two related tables are both selected.

\subsection{LLM-based Data Product Creation}

Given a JSON-formatted schema and a DPR, we prompt the LLM with chain-of-thought reasoning to (1) identify relevant columns and (2) generate derived columns with SQL transformations. This is more expensive than hybrid search but uniquely enables derived column generation. We evaluate 10 LLMs spanning open-weight models and frontier closed-weight systems (details in Appendix \ref{sec:appendix}).


\section{Experiment Results}

\subsection{Column Selection Results}
First, we highlight key results for column-level metrics for automatic data product creation using our baseline approaches of hybrid search and LLM-based DP creation using 10 different LLMs spanning open-weight and closed-weight models. Detailed result tables for each experiment can be found in Appendix \ref{apen-column-results}.

\begin{figure}[h]
    \centering
    \includegraphics[width=\linewidth]{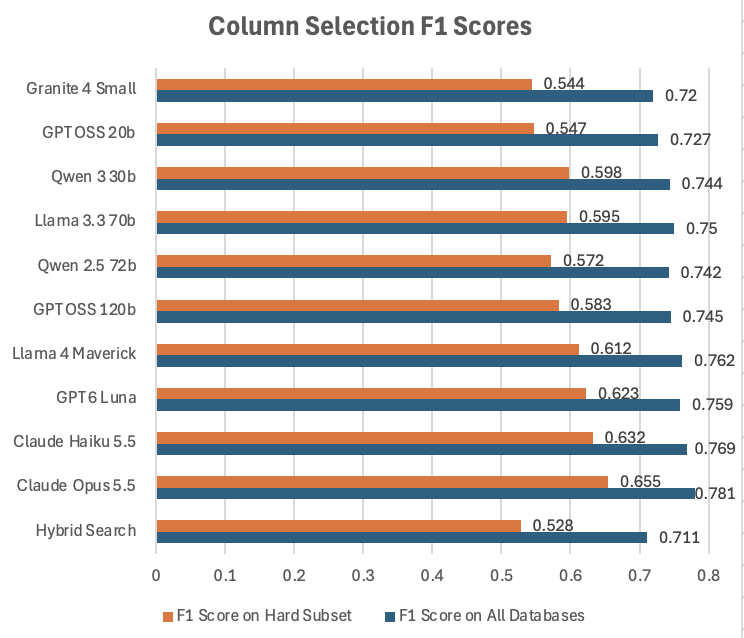}
    \caption{F1 scores for data product column selection across methods, comparing results on all DP-Bench databases and the Hard subset.}
    \label{fig:col_selection_summary}
\end{figure}

Figure \ref{fig:col_selection_summary} shows baseline results over all databases and DPRs in DP-bench, as well as over only the \textit{hard} subset of databases in DP-bench, which consists of 30 databases which contain over 50 total columns. This chart highlights results for only F1 scores and considers all provenance columns for derived columns as also being part of the DP. When looking only at F1, it appears that most of our LLM-based approaches show very similar performance, although in our full results, we can also observe some differences in performance for each model (e.g., both GPT OSS models tended towards having higher recall). Notably, the three frontier closed-weight models---Claude Opus 5.5, Claude Haiku 5.5, and GPT 6 Luna---achieve the highest F1 scores overall, with Claude Opus 5.5 reaching the top F1 on all databases.

We also can see that the hard subset, which focuses on databases with more columns, shows a decrease in F1 of roughly 0.15--0.2 across all models. We generally can see that the absolute scores for the Hard subset are lower across the board, supporting the increased challenge of the Hard subset.

\subsubsection{Derived Column Results}

\begin{figure}[h]
    \centering
    \includegraphics[width=\linewidth]{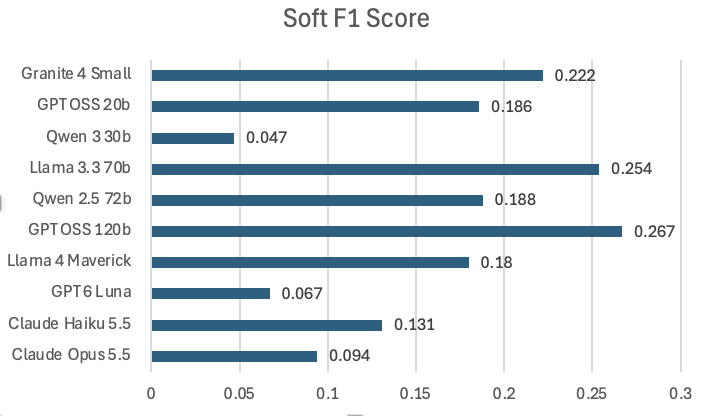}
    \caption{Soft-F1 scores of derived column predictions.}
    \label{fig:derived_col_sim}
\end{figure}

Figure \ref{fig:derived_col_sim} shows the computed soft-F1 scores for the derived columns predicted by our baseline approaches. Full result tables can be found in Appendix \ref{apen-column-results}. Note that hybrid search is not included as a baseline because it does not produce any derived columns. 

Compared to the standard column selection, we see greater variation in performance across the different LLMs used for this task. For example, while Qwen 3 30b was one of the best performing models for basic column selection, it struggled significantly in producing derived columns which were similar to the ground truth. On the other hand, Granite 4 showed relatively stronger performance at producing derived columns compared to its performance on column selection.
Notably, the best derived column results come from open-weight models: GPT OSS 120b achieves the highest soft-F1 (0.267), followed by Llama 3.3 70b (0.254). Among the three frontier closed-weight models, Claude Haiku 5.5 (0.131) outperforms Claude Opus 5.5 (0.094) and GPT 6 Luna (0.067), suggesting that larger frontier models do not automatically excel at the harder sub-task of transformation generation.
Producing derived columns which closely match those in the ground-truth DP appears to be a much more challenging task across all databases, as evidenced by the fact that performance on the Hard subset of DP-Bench shows nearly identical results as on all databases.

\subsection{Text-to-SQL Evaluation}

Lastly, we present results for the downstream text-to-SQL task over DP-Bench. To perform this evaluation, we consider 2 scenarios: first, the \textit{Oracle} setting, where we assume the use of a perfect text-to-SQL translation system, and evaluate the highest possible execution accuracy that could be achieved for a given DP; second, using Llama 4 Maverick to perform text-to-SQL generation.

Execution accuracy for our text-to-SQL task is shown in Table \ref{tab:res_execacc}, where we perform tests only for questions contained in the Hard subset of DP-Bench (900 total questions across 30 databases). In addition to our baseline DP creation methods, we also present results using ``no search'' (i.e. use the entire database) and the ground-truth DP as input. Importantly, we observe that for text-to-SQL, using the ground-truth DP leads to better performance than using the entire database, supporting our underlying assumption that selecting a subset of the database as the DP helps to provide a more focused schema for downstream tasks and reduces errors.

A notable pattern emerges when comparing the frontier closed-weight models to the open-weight baselines. The three frontier models (Claude Opus 5.5, Claude Haiku 5.5, GPT 6 Luna) all achieve Oracle scores of 90--91\%, substantially higher than the best open-weight model, GPT OSS 120b, at 84.3\%. This indicates that frontier models produce DPs with significantly better table coverage. However, their Llama 4 Maverick text-to-SQL accuracy is lower (29.7--30.7\%) than GPT OSS 120b (37.6\%). These results suggest that factors beyond table coverage may influence downstream text-to-SQL performance.

\begin{table}[h]
    \centering
    \small
    \begin{tabular}{l c c c}
    Method & Oracle & Llama 4 Mav. & Input Tokens\\
    \hline
    No Search & 100\% & 45.6 \%  & 8,432 \\
    Ground-Truth & 100\% & 49.8 \% & 2,192 \\
    \hline \hline
    HybridSearch & 63.0\% &  31.3\%  & 6,186 \\
    \hline
    \hline
    \multicolumn{4}{c}{\tt Frontier Models}\\
    \hline
    Claude Opus 5.5 &	90.0\%	& 30.7\% &	3,498 \\
    Claude Haiku 5.5 &	\bf 91.3\% & 30.7\% &	3,811 \\
    GPT 6 Luna &	90.7\% &	29.7\% &	3,791 \\
    \hline
    \multicolumn{4}{c}{\tt Other Models}\\
    \hline
    Llama 4 Maverick & 43.4\% & 22.2\%  & \bf 2,218 \\
    GPT OSS 120b & 84.3\% & \bf 37.6\%  & 3,587 \\
    Qwen 2.5 72b & 63.6\% & 31.2\%  & 3,031 \\
    Llama 3.3 70b & 58.4\% & 28.1\%  & 2,780 \\
    Qwen 3 30b & 47.0\% & 24.4\%  & 2,405 \\
    GPT 20b & 80.1\% & 36.3\%  & 3,630 \\
    Granite 4 small & 53.7\% & 26.0\%  & 2,730 \\
    \end{tabular}
    \caption{Text-to-SQL execution accuracy over the Hard subset. Oracle indicates the best possible accuracy over the DP, compared to using Llama 4 for Text-to-SQL.}
    \label{tab:res_execacc}
\end{table}

In general, text-to-SQL results correspond to recall: without the necessary columns in the DP, questions cannot be answered regardless of other quality aspects. Frontier models balance high Oracle (90--91\%) with token efficiency (3,498--3,811 tokens), comparable to open-weight baselines. Producing a precise DP schema still matters: the rightmost column of Table \ref{tab:res_execacc} shows that high-recall but low-precision approaches (e.g., No Search at 8,432 tokens) incur substantially higher inference cost. For example, the largest database in DP-Bench contains 455 columns while its ground-truth DP has only 30. Without performing search, the average input token length already exceeds the context limit (8,192 tokens) of many older models, making effective data product creation crucial for reducing schema size while retaining downstream accuracy.

\section{Key Findings and Insights}

We find that column selection can be relatively well handled by baseline approaches, while derived column generation remains a major challenge. Recall is critical for downstream task performance, and using the full database can outperform filtered data products---the usefulness of data products thus depends heavily on the precision--recall tradeoff. Frontier closed-weight models achieve the best column selection F1 and the highest Oracle scores in text-to-SQL (90--91\%), yet their realized execution accuracy falls short of the best open-weight models, suggesting that DP quality alone does not determine downstream performance.

{\small
\begin{itemize}[nosep, leftmargin=7pt]
    \item Column selection is relatively well-handled by current approaches.
    \item Derived column generation remains a major challenge.
    \item Recall is critical for downstream task performance.
    \item Data product usefulness depends on the precision--recall trade-off.
\end{itemize}
}
Further details are in Appendix \ref{apen-insights}.

\section{Related Work}
Prior work on data products includes overviews~\cite{hasan2023} and agentic optimization~\cite{tamilselvan2025}, but none propose automatic generation or a benchmark. \citet{zhang2025} builds a data product benchmark from HybridQA via clustering, but lacks manual vetting, derived columns, topic annotations, and baseline methods—all present in DP-Bench. TARGET~\cite{ji2024target} retrieves tables/columns for factoid queries rather than high-level business abstractions. \cite{DAmbrosioPSSRR25} detects catalog duplications but does not generate DPs. Schema linking~\cite{lei-etal-2020-examining} produces per-query schema reductions, not reusable DPs with derived columns.

\section{Conclusion}
We introduced \textbf{DP-Bench}, a first-of-a-kind benchmark for automatic data product creation, with manually validated DPRs, data products with base and derived columns, SQL provenance, and baseline approaches. Results show that column selection is tractable while derived column generation remains challenging, and that balancing precision and recall is critical for downstream performance. DP-Bench opens research directions in complex schema handling, transformation generation, cross-database DPs, and agent-based optimization.

\section*{Limitations}


\subsection*{Boundaries and Assumptions}

In this paper, some boundary decisions and assumptions regarding the scope of our work lead to limitations regarding the coverage of what aspects of Data Products we address and evaluate.

To enable objective evaluation and tractability of the task, we impose the following boundaries on our work in this paper:

\begin{itemize}
    \item \textbf{Relational focus:} We consider structured relational databases as the primary data source, with optional inclusion of unstructured text.
    \item \textbf{Schema-level outputs:} Data products are represented as collections of tables and columns, rather than full-fledged artifacts such as dashboards, reports, or applications.
    \item \textbf{Single-request setting:} Each data product is generated independently for a single DPR, without iterative refinement or feedback loops.
    \item \textbf{Static generation:} We focus on the initial generation of a data product, rather than its lifecycle management or evolution over time.
    \item \textbf{Non-uniqueness:} Multiple valid data products may satisfy the same DPR; therefore, evaluation is based on similarity rather than exact matching.
\end{itemize}

Several important aspects of data product creation, usage, and maintenance are outside the scope of this work:
\begin{itemize}
    \item Visualization layers such as dashboards and reports,
    \item Cross-database or distributed data product construction,
    \item Continuous optimization or monitoring of data products,
    \item Data quality assessment and error remediation.
\end{itemize}


\bibliography{custom}

\appendix

\section{Appendix}

\section{Models used for experiments}
In our experiments, we investigate a wide range of LLM sizes and architectures, including MoE models (GPT OSS 20b and GPT OSS 120b \cite{openai2025gptoss120bgptoss20bmodel}, Qwen 3 30b \cite{yang2025qwen3technicalreport}, Llama 4 Maverick \cite{llama4}), a hybrid transformer-mamba architecture (Granite 4 Small \cite{granite4}), and dense open-weights models (Qwen 2.5 72b \cite{qwen2025qwen25technicalreport}, Llama 3.3 70b \cite{grattafiori2024llama3herdmodels}). 

To ensure comprehensive coverage across deployment modalities, we also evaluate leading closed-weight proprietary systems across both lightweight tiers (Claude Haiku 5.5 \cite{anthropic2026haiku}, GPT 6 Luna \cite{openai2026gpt6luna}) and frontier reasoning tiers (Claude Opus 5.5 \cite{anthropic2026opus}).

\label{sec:appendix}

\subsection{Prompt to generate provenance in SQL for the derived columns}
\label{apen-provenance}

We used the following prompt to generate provenance in SQL for the derived columns:
\begin{quote}
    \tt Generate only SQL to find \{COL\_DESC\} from the following database schema.
    \\~\\
    \{DB\}
\end{quote}
~\\
Here \emph{``\{COL\_DESC\}''} is the description of the derived columns (obtained from the corresponding ELT-Bench data model) and \emph{``\{DB\}''} is the complete database schema for the corresponding BIRD database. Such a schema information includes table names, column names, column descriptions, data type of the columns, the primary keys for each table and the foreign keys for each table.

\subsection{Prompt to generate DP summaries}
\label{apen-dp-summary}

Given the tables of a data product, we used the LLM to generate 3 possible summaries using the following prompt -

\begin{quote}
    \tt Summarize the following tables.
    \\~\\
    \{DP\_TABLES\}
    \\~\\
    The summary should be limited to 100 words. Try to avoid numbers. The summarization should start with something like ``This table contains data that enables analysis of ...''
\end{quote}

Here, \emph{``\{DP\_TABLES\}''} is a placeholder for the metadata of the tables in a data product. Such metadata includes table names, column names (both derived and non-derived) and the column descriptions.

\subsection{Prompt to generate candidate data product requests}
\label{apen-dpr}

Given a summary, we used the LLM to convert it into a \textbf{candidate data product request} using the following prompt containing an example --

\begin{quote}
    \tt Transform the input text using the following example. Do not add any explanation.
\\~\\
Input Text: This table contains data that enables analysis of geographic and demographic information, including population characteristics,
 economic data, and congressional representation. It covers various aspects of residential areas, such as household income, education, and 
 employment, as well as information about the location itself, like elevation and time zone. The data also includes details about the local 
 government and organization, allowing for a comprehensive understanding of the area.
\\~\\
Transformed text: Collect data that will allow queries on geographic and demographic information, including population characteristics,
 economic data, and congressional representation. It should cover various aspects of residential areas, such as household income, education, and 
 employment, as well as information about the location itself, like elevation and time zone. It should also include details about the local government and organization, allowing for a comprehensive understanding of the area.
\\~\\
Input Text: \{SUMMARY\}
\\~\\
Transformed text:
\end{quote}
~\\

Here \emph{``\{SUMMARY\}''} refers to a summary of a data product generated earlier. 

\subsection{Derived Columns Filtering}
\label{app:der_col_fil}
During this step, we had to filter out some derived columns due to erroneous provenance. For example, consider the example shown in Figure \ref{fig:example_bad_derived_col} in the DB \texttt{movielens}. The column description from the corresponding ELT-Bench data model is confusing, and as a result the generated SQL as provenance is actually useless. So, we discarded it. Finally, after the aforementioned filtering, we have  \textbf{582} derived columns for consideration.

\begin{figure}[h]
\scriptsize
\begin{mdframed}
\emph{``column name''}: ``is\_called\_box\_office\_success\_paradox''\\
\emph{``description''}: ``Set to 1 if the director is called the box office success paradox, otherwise, set to 0''\\
\emph{``generated provenance''}: ``SELECT CASE          WHEN directorname = 'The Box Office Success Paradox' THEN 1          ELSE 0      END  FROM      directors;''
\end{mdframed}
\caption{Example of confusing derived column.}
\label{fig:example_bad_derived_col}
\end{figure}

\subsection{Topic annotation}
\label{app:topic}
To generate topics for each DP, we employed the Llama-3.3 70B model using manually curated three-shot prompting examples. The generated topics were then manually reviewed by annotators to remove irrelevant or redundant entries that did not accurately reflect the tables and columns within the corresponding DP. This process resulted in a total of \textbf{564} topics across the 78 data products.

For instance, for a representative DP (see Figure \ref{fig:example_dp}), the annotated topics include: \textit{User reviews, Mobile apps, Android apps, App reviews, App ratings, App installation, Sentiment analysis,} and \textit{Play store}.

We applied the same topic generation and validation procedure to DPRs associated with each DP. This resulted in a separate set of \textbf{461} topics derived from DPRs.

Continuing the earlier example (Figure \ref{fig:example_dp_input}), the topics generated from the corresponding DPR include: \textit{User reviews, App performance evaluation, App downloads, Sentiment analysis,} and \textit{Mobile app reputation}. 

\subsection{DP quality evaluation using topics}
\label{app:topic-exp}

\begin{figure}[h]
  \centering
  \includegraphics[width=0.5\textwidth]{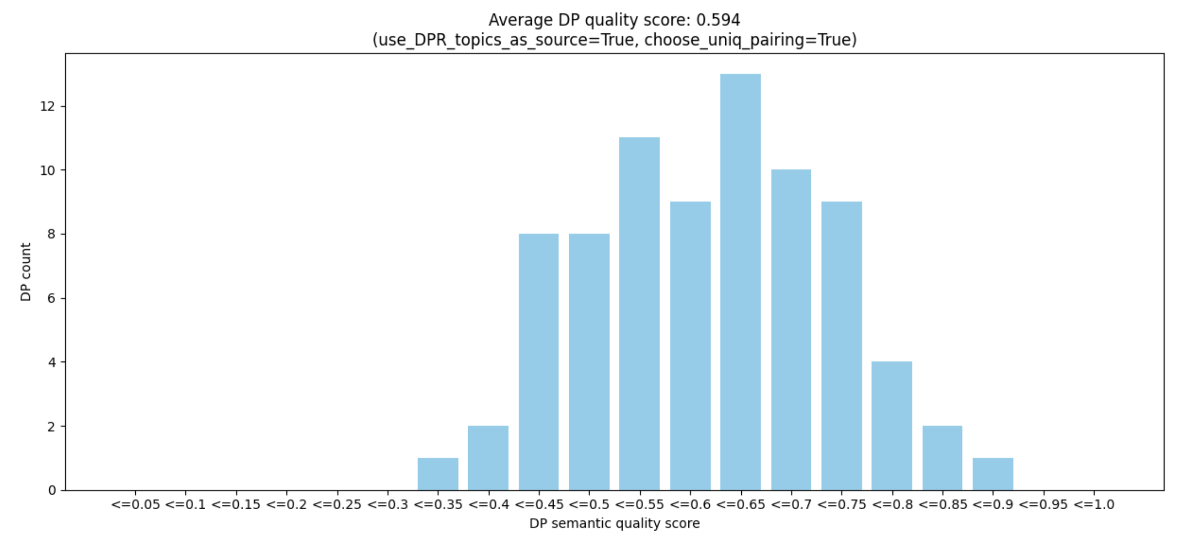}
  \caption{\centering Distribution of semantic quality scores for the generated DPs from the \textit{GPT OSS 120b} based method.}
  \label{fig:dp_quality_gptloss}
\end{figure}

Using the approach described in Section \ref{subsec:anno_topics}, we generated topics for the generated DPs. For each topic from a gold DPR, we selected the best possible topic from the topics of the corresponding generated DP by computing a similarity score between the embeddings using the \textit{IBM slate.30m.english.rtrvr} model. The average of the best similarity scores for the selected topics of the corresponding generated DP is used as the semantic quality score for the generated DP. 

Figure \ref{fig:dp_quality_gptloss} shows the distribution of topic-based semantic quality score of the generated DPs (for each of the 78 DBs) from our best baseline method of \textit{GPT OSS 120b}. The average DP quality score is \emph{0.594}. Our analysis indicates that there is no correlation between the size of the original DB and the eventual topic-based quality score for the corresponding generated DP given the DPR. From our random manual analysis, it appears that usually the more vague or less precise the DPR (i.e. the business use case) is, the lower the generated DP quality score.

\subsection{Expanded Key Findings and Insights}
\label{apen-insights}

In this section, we summarize the key insights obtained from our experimental evaluation of automatic data product creation systems on DP-Bench. These findings highlight both the progress made by current approaches and the remaining challenges in this problem space.

\textbf{Insight 1: Column selection is relatively well-handled by current approaches.}
Across all evaluated methods, particularly LLM-based approaches, we observe reasonably strong performance in selecting relevant columns, with F1 scores reaching close to 0.7 on the full benchmark. This indicates that modern models are effective at identifying relevant subsets of the schema corresponding to a given DPR, especially when compared to naive baselines.

\textbf{Insight 2: Derived column generation remains a major challenge.}
In contrast to column selection, generating derived columns with correct transformations proves significantly more difficult. All evaluated methods achieve relatively low soft-F1 scores on derived column similarity, suggesting that accurately modelling transformations and feature engineering logic remains an open problem.

\textbf{Insight 3: Recall is critical for downstream task performance.}
Our analysis of the text-to-SQL evaluation reveals a strong correlation between recall in column selection and execution accuracy. Systems that fail to include necessary columns in the generated data product cannot support correct query generation, even if other aspects of the data product are accurate.

\textbf{Insight 4: Using the full database can outperform filtered data products.}
Surprisingly, we observe that in some cases, using the entire database (i.e., the no-search baseline) yields higher text-to-SQL execution accuracy than generated data products. This finding suggests that modern LLMs are increasingly capable of handling large schemas, and that aggressive schema pruning can negatively impact performance when recall is insufficient.

\textbf{Insight 5: Data product usefulness depends on the precision--recall trade-off.}
While full-schema approaches may achieve higher recall, they come at the cost of significantly larger input sizes and increased computational overhead. In contrast, data products provide a more compact and focused schema, reducing prompt size and improving efficiency. This highlights the importance of balancing precision and recall in data product creation.

\textbf{Insight 6: The quality of DPRs strongly influences outcomes.}
Our analysis indicates that the clarity and specificity of data product requests play a crucial role in determining the quality of generated data products. Vague or underspecified DPRs tend to produce lower-quality outputs, as reflected in semantic similarity and topic-based evaluations.

\textbf{Insight 7: Benchmark difficulty reveals performance gaps.}
Results on the Hard subset of DP-Bench, consisting of larger and more complex databases, show consistently lower performance across all methods. This demonstrates that current approaches struggle to scale effectively with schema complexity, highlighting an important direction for future work.

Overall, our findings show that while current methods are capable of identifying relevant schema elements, generating high-quality, transformation-rich data products remains challenging. Furthermore, the interplay between schema reduction and downstream task performance suggests that more principled approaches are needed to balance efficiency and effectiveness in automatic data product creation.

\subsection{Detailed Results}
\label{apen-column-results}

Here, we show detailed results for experiments surrounding column selection.

\subsubsection{Column Selection Precision, Recall, and F1}

Table \ref{tab:res_prf1} shows baseline results over all databases and DPRs in DP-Bench. We present results in terms of computing precision, recall, and F1 on \textit{non-derived} columns only, where we ignore all derived column content from the predicted and ground-truth DP. Similarly, Table \ref{tab:res_prf1_id} shows results for the \textit{incl. derived} setting, where we consider all provenance columns for derived columns as also being part of the DP, and compute each metric accordingly. Additionally, we indicate precision, recall, and F1 metrics for a \textit{No Search} baseline, where we simply use the entire database as the Data Product and compute metrics.

\begin{table}[h]
    \centering
    \begin{tabular}{l | c c c }
    Method & P & R & F1  \\
    \hline
    No Search & 0.522 & 1.00 & 0.648 \\
    \hdashline
    Hybrid Search & 0.575 & 0.872 & 0.652 \\
    \hline
    \multicolumn{4}{c}{\tt Frontier Models}\\
    \hline
    Claude Opus 5.5 & \bf 0.664 & 0.848 & \bf 0.725 \\
    Claude Haiku 5.5 & 0.624 & 0.857 & 0.702 \\
    GPT 6 Luna &  0.632 & 0.840 & 0.697 \\
    \hline
    \multicolumn{4}{c}{\tt Other Models}\\
    \hline
    Llama 4 Maverick & 0.654 & 0.807 & 0.704 \\
    GPT OSS 120b & 0.563 & \textbf{0.952} & 0.677  \\
    Qwen 2.5 72b & 0.582 & 0.890 & 0.679  \\
    Llama 3.3 70b & 0.601 & 0.880 & 0.693  \\
    Qwen 3 30b & 0.637 & 0.802 & 0.687 \\
    GPT OSS 20b & 0.558 & 0.931 & 0.662  \\
    Granite 4 Small & 0.590 & 0.835 & 0.665  \\
    \end{tabular}
    \caption{Precision (P), Recall (R), and F1 scores for DP creation using baseline approaches computed over only non-derived columns.}
    \label{tab:res_prf1}
\end{table}

\begin{table}[h]
    \centering
    \begin{tabular}{l | c c c}
    Method & P & R & F1 \\
    \hline
    No Search & .598 & 1.00 & .715 \\
    \hdashline
    Hybrid Search &  0.653 & 0.869 & 0.711 \\
    \hline
    \multicolumn{4}{c}{\tt Frontier Models}\\
    \hline
    Claude Opus 5.5 & \bf 0.758 & 0.837 & \bf 0.781 \\
    Claude Haiku 5.5 & 0.722 & 0.858 & 0.769 \\
    GPT 6 Luna &  0.725 & 0.835 & 0.759 \\
    \hline
    \multicolumn{4}{c}{\tt Other Models}\\
    \hline
    Llama 4 Maverick & 0.749 & 0.802 & 0.762 \\
    GPT OSS 120b & 0.646 & \textbf{0.950} & 0.745 \\
    Qwen 2.5 72b &  0.665 & 0.887 & 0.742 \\
    Llama 3.3 70b & 0.688 & 0.871 & 0.750 \\
    Qwen 3 30b &  0.728 & 0.797 & 0.744 \\
    GPT OSS 20b &  0.638 & 0.929 & 0.727 \\
    Granite 4 Small &  0.670 & 0.829 & 0.720 \\
    \end{tabular}
    \caption{Precision (P), Recall (R), and F1 scores for DP creation using baseline approaches while including the source columns used to produce derived columns.}
    \label{tab:res_prf1_id}
\end{table}

We can observe that even without performing any search, an F1 score of .648 can be achieved due to our benchmark containing many small databases, where simply selecting all columns in the database results in relatively high precision. However, we do still observe that all baselines out-perform the no search baseline in terms of precision and F1 score with the exception of Hybrid Search in the Incl. Derived setting.

Across the LLM-based approaches, we also can observe some key differences in performance for each metric. Notably, the GPT OSS models both tend towards having very high recall at the cost of lower precision. Among frontier models, Claude Opus 5.5 achieves the highest F1 in both settings. Among open-weight models, Llama 4 Maverick and Llama 3.3 70b show the strongest F1 scores.

\begin{table}[h]
    \centering
    \begin{tabular}{l | c c c }
    Method & P & R & F1 \\
    \hline
    No Search & 0.310 & 1.00 & 0.459  \\ \hline
    Hybrid Search & 0.358 & 0.819 & 0.470  \\ \hline
    \hline
    \multicolumn{4}{c}{\tt Frontier Models}\\
    \hline
    Claude Opus 5.5 & 0.518 & 0.757 & \bf 0.597 \\
    Claude Haiku 5.5 & 0.470 & 0.761 & 0.561 \\
    GPT 6 Luna & 0.477 & 0.750 & 0.561  \\
    \hline
    \multicolumn{4}{c}{\tt Other Models}\\
    \hline
    Llama 4 Maverick & \textbf{0.520} & 0.649 & 0.557 \\
    GPT OSS 120b & 0.371 & \textbf{0.905} & 0.507 \\
    Qwen 2.5 72b & 0.396 & 0.779 & 0.504  \\
    Llama 3.3 70b & 0.443 & 0.757 & 0.537  \\
    Qwen 3 30b & 0.476 & 0.675 & 0.534  \\
    GPT OSS 20b & 0.363 & 0.872 & 0.480 \\
    Granite 4 Small & 0.409 & 0.697 & 0.489  \\
    \end{tabular}
    \caption{Precision (P), Recall (R), and F1 scores for baseline approaches over the Hard subset (30 databases which contain over 50 total columns) of our benchmark, computed only for non-derived columns.}
    \label{tab:res_prf1_hard}
\end{table}

\begin{table}[h]
    \centering
    \begin{tabular}{l | c c c }
    Method & P & R & F1 \\
    \hline
    No Search & 0.381 & 1.00 & 0.534 \\ \hline
    Hybrid Search &  0.429 & 0.805 & 0.528 \\ \hline
    \hline
    \multicolumn{4}{c}{\tt Frontier Models}\\
    \hline
    Claude Opus 5.5 & 0.616 & 0.742 & \bf 0.655 \\
    Claude Haiku 5.5 & 0.571 & 0.756 & 0.632 \\
    GPT 6 Luna & 0.571 & 0.740 & 0.623 \\
    \hline
    \multicolumn{4}{c}{\tt Other Models}\\
    \hline
    Llama 4 Maverick & \textbf{0.619} & 0.641 & 0.612 \\
    GPT OSS 120b & 0.454 & \textbf{0.903} & 0.583 \\
    Qwen 2.5 72b &  0.479 & 0.777 & 0.572 \\
    Llama 3.3 70b &  0.532 & 0.742 & 0.595 \\
    Qwen 3 30b & 0.578 & 0.670 & 0.598 \\
    GPT OSS 20b &  0.437 & 0.866 & 0.547 \\
    Granite 4 Small & 0.490 & 0.690 & 0.544 \\
    \end{tabular}
    \caption{Precision (P), Recall (R), and F1 scores for baseline approaches over the Hard subset including derived columns.}
    \label{tab:res_prf1_hard_id}
\end{table}

Table \ref{tab:res_prf1_hard} and Table \ref{tab:res_prf1_hard_id} show the same metrics over only the \textit{hard} subset of databases in DP-bench, which consist of 30 databases which contain over 50 total columns. Here, the value of generating data products is greater, as the no search baseline only has a precision of .310, indicating that a much larger number of columns are not useful for down-stream tasks. The relative performance among approaches appears to be the same for this subset, but we observe that the best baseline -- using Llama 4 Maverick -- improves on the no search and hybrid search baselines by nearly 0.1. We generally can see that the absolute scores for the Hard subset are lower across the board, supporting the increased challenge of the Hard subset.

\subsubsection{Derived Column Similarity Metrics}
Next, we present results for derived column similarity metrics, computed as soft-precision, soft-recall, and soft-F1 scores. Table \ref{tab:res_deriveds} shows results over all databases in DP-Bench, and Table \ref{tab:res_deriveds_hard} presents results for only the hard subset of databases. Note that hybrid search is not included as a baseline because it does not produce any derived columns. 

\begin{table}[h]
    \centering
    \begin{tabular}{l c c c }
    Method & sP & sR & sF1 \\ 
    \hline
    \hline
    \multicolumn{4}{c}{\tt Frontier Models}\\
    \hline
    Claude Opus 5.5 &  0.186 &	0.070 &	0.094  \\
    Claude Haiku 5.5 & 0.192 &	0.109 &	0.131  \\
    GPT 6 Luna & 0.141 &	0.050 &	0.067 \\
    \hline
    \multicolumn{4}{c}{\tt Other Models}\\
    \hline
    Llama 4 Maverick & 0.298 & 0.135 & 0.180 \\
    GPT OSS 120b & 0.270 & \textbf{0.290} & \textbf{0.267} \\
    Qwen 2.5 72b & 0.250 & 0.165 & 0.188 \\
    Llama 3.3 70b & \textbf{0.329} & 0.221 & 0.254 \\
    Qwen 3 30b & 0.101 & 0.034 & 0.047 \\
    GPT OSS 20b & 0.176 & 0.241 & 0.186 \\
    Granite 4 Small & 0.290 & 0.195 & 0.222 \\
    \end{tabular}
    \caption{Soft-precision, -recall, and -F1 scores of the derived columns produced by each baseline method.}
    \label{tab:res_deriveds}
\end{table}

\begin{table}[h]
    \centering
    \begin{tabular}{l c c c }
    Method & sP & sR & sF1 \\ 
    \hline
    \multicolumn{4}{c}{\tt Frontier Models}\\
    \hline
    Claude Opus 5.5 &  0.176 &	0.069 &	0.093  \\
    Claude Haiku 5.5 & 0.186 &	0.109 & 0.130  \\
    GPT 6 Luna & 0.129 &	0.042 &	0.059 \\
    \hline
    \multicolumn{4}{c}{\tt Other Models}\\
    \hline
    Llama 4 Maverick & 0.299 & 0.136 & 0.180 \\
    GPT OSS 120b & 0.273 & \textbf{0.282} & \textbf{0.266} \\
    Qwen 2.5 72b & 0.245 & 0.156 & 0.179 \\
    Llama 3.3 70b & \textbf{0.338} & 0.218 & 0.254 \\
    Qwen 3 30b & 0.100 & 0.038 & 0.050 \\
    GPT OSS 20b & 0.177 & 0.235 & 0.185 \\
    Granite 4 Small & 0.305 & 0.193 & 0.223 \\
    \end{tabular}
    \caption{Soft-precision, -recall, and -F1 scores of derived columns over the Hard subset tables.}
    \label{tab:res_deriveds_hard}
\end{table}

Compared to the standard column selection, we see some greater variation in performance across the different LLMs used for this task. For example, while Qwen 3 30b was one of the best performing models for basic column selection, it struggled significantly in producing derived columns which were similar to the ground truth. On the other hand, Granite 4 showed relatively stronger performance at producing derived columns compared to its performance on column selection.

Producing derived columns which closely match those in the ground-truth DP appears to be a much more challenging task across all databases, as evidenced by the fact that performance on the Hard subset of DP-Bench shows nearly identical results as on all databases.

\subsection{Statistical details of the data products in DP-Bench}
\label{apen-dp-details}

\begin{figure*}[h]
  \centering
  \includegraphics[width=1.0\textwidth]{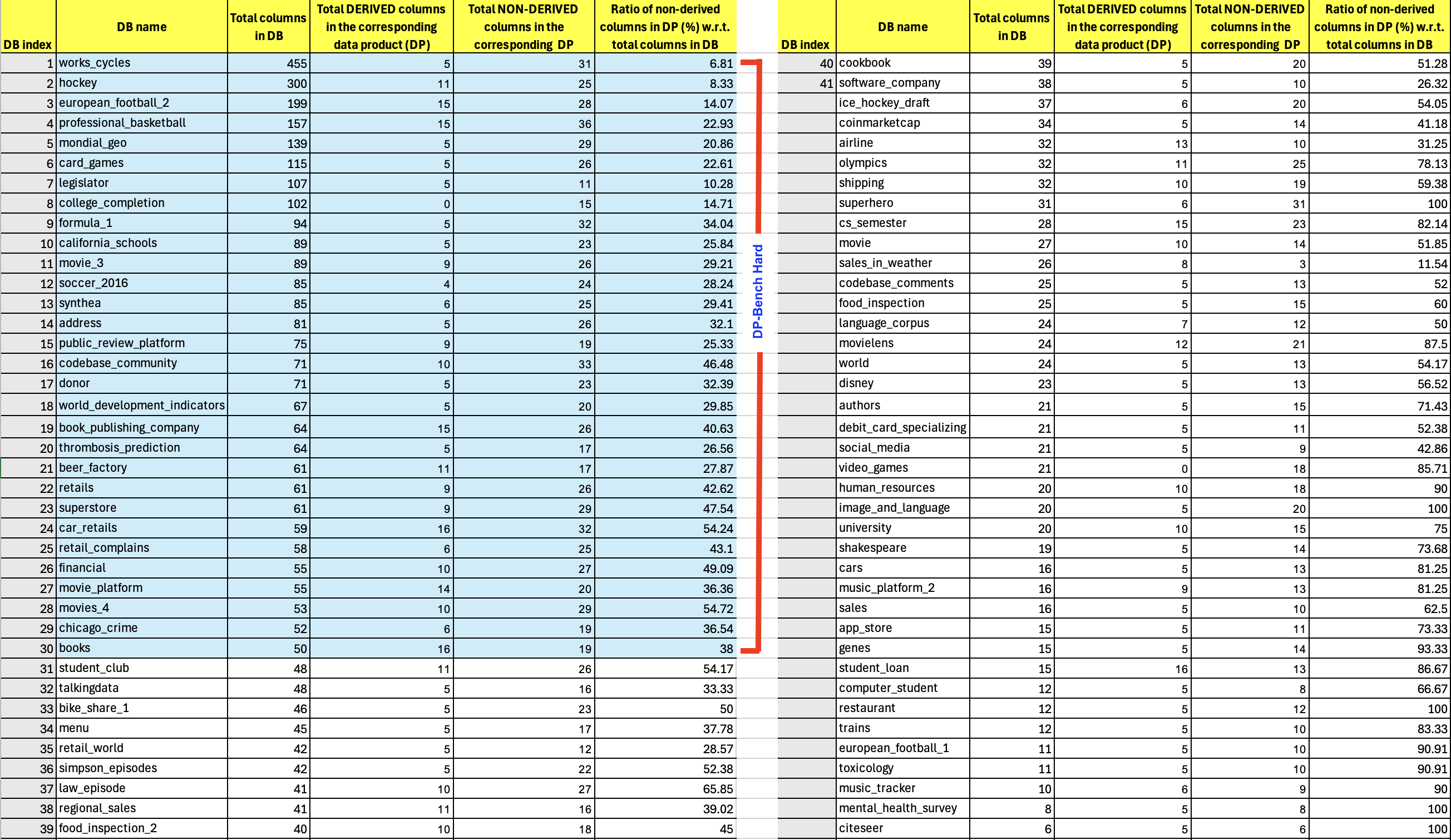}
  \caption{\centering The number of derived and non-derived columns in the data product for each DB in the DP-Bench.}
  \label{fig:dp_bench_dist}
\end{figure*}

Figure \ref{fig:dp_bench_dist} shows various statistics including number of derived and non-derived columns in the data products in DP-Bench. The first 30 DBs are part of the DP-Bench Hard subset.

\subsection{Examples from the BIRD and ELT-Bench benchmarks}
\label{apen-bird-elt-bench}

\begin{figure}[h]
  \centering
  \includegraphics[width=0.5\textwidth]{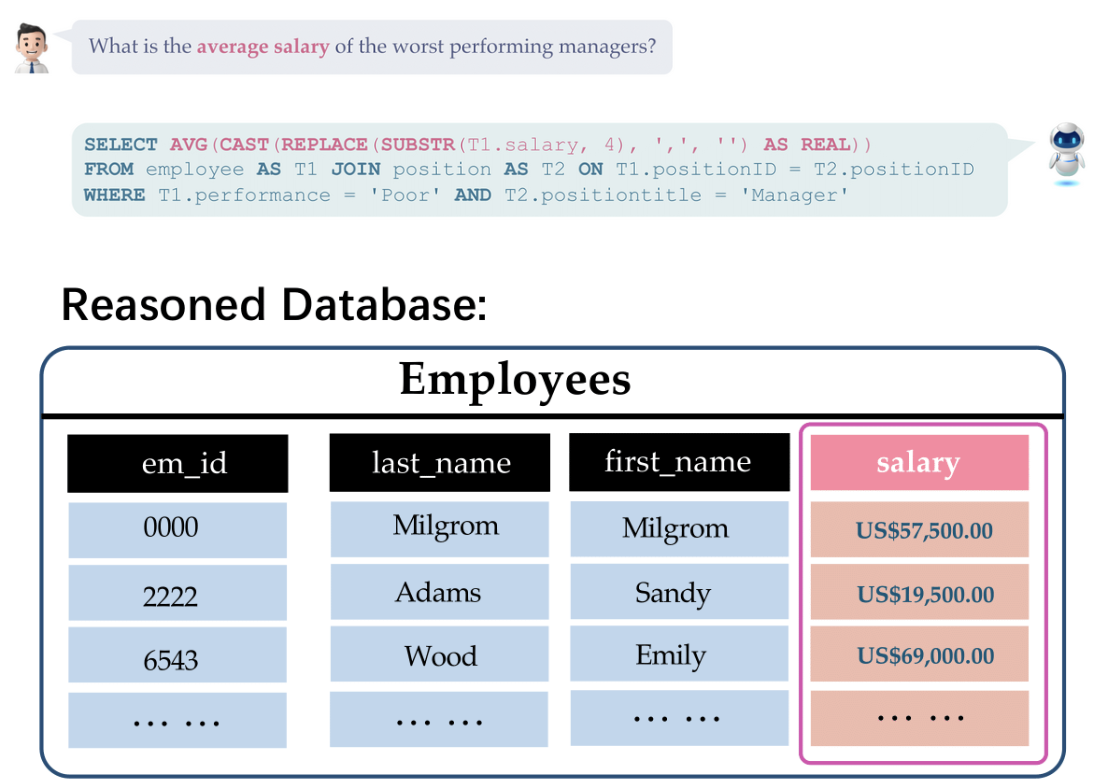}
  \caption{\centering Example of a natural question--SQL pair in BIRD from \url{https://bird-bench.github.io}}
  \label{fig:example_bird}
\end{figure}

\begin{figure}[h]
  \centering
  \includegraphics[width=0.5\textwidth]{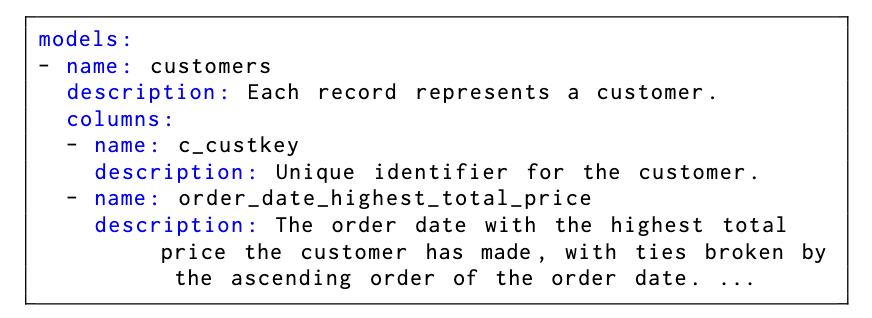}
  \includegraphics[width=0.5\textwidth]{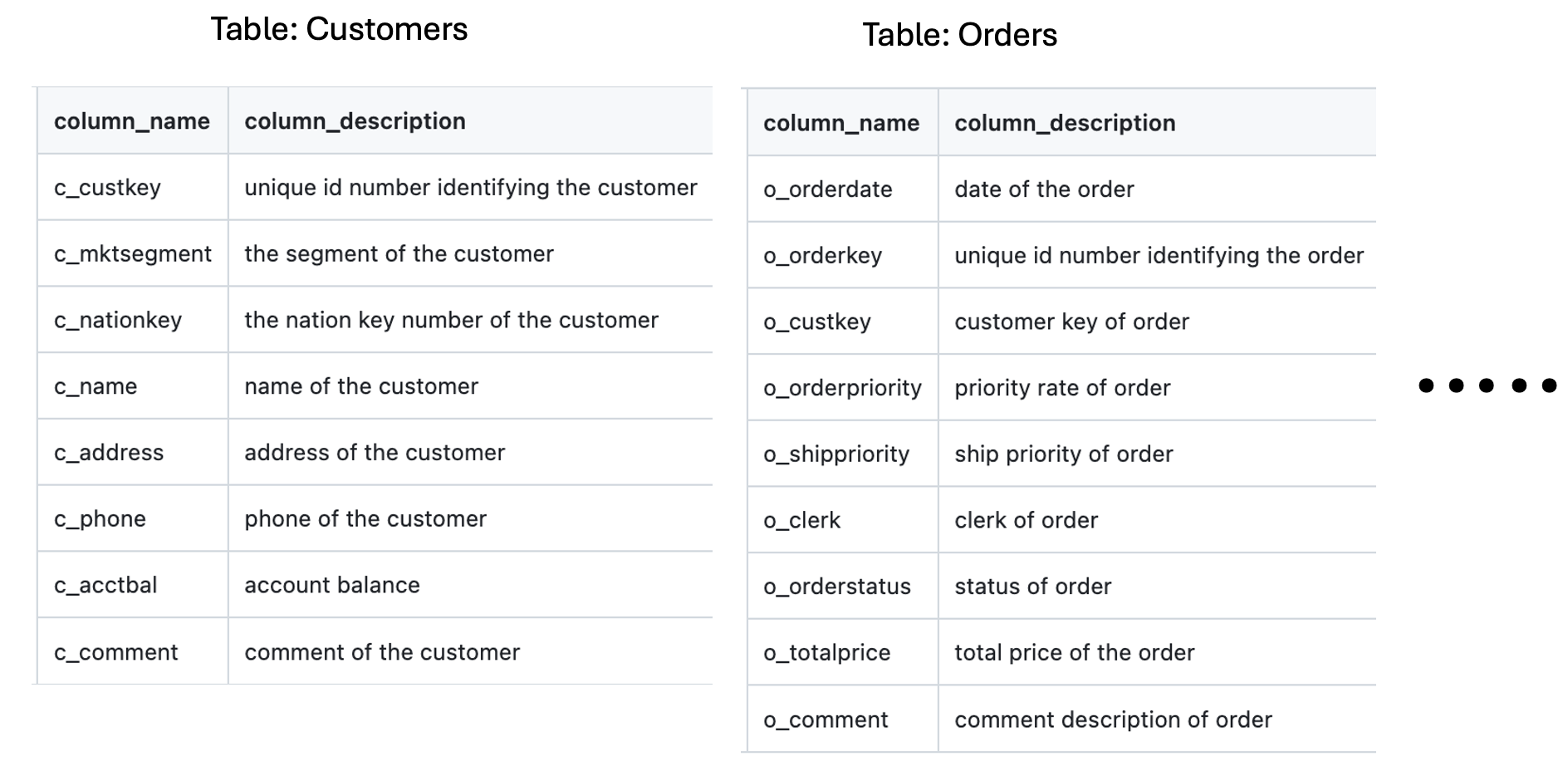}
  \caption{\centering A partial description of the ``customers'' data model for the ``retails'' database in ELT-Bench.}
  \label{fig:example_elt_bench}
\end{figure}

Figure \ref{fig:example_elt_bench} shows an example of a data model named ``customers'' that is created from the database ``retails''. The data model column ``c\_custkey'' can be directly traced to the a column with the same name inside the table ``Customers'' in the ``retails'' database. On the other hand, the data model column ``order\_date\_highest\_total\_price'' is derived by consolidating information from several columns in tables ``Customers'' and ``Orders'' in that database. 

\subsection{License For Artifacts and Intended Use}

The license for DP-Bench is CC BY-NC-SA 4.0 (Creative Commons Attribution-NonCommercial-ShareAlike 4.0 International). BIRD dataset has CC BY-SA 4.0 license. ELT-Bench has CC BY-NC-ND 4.0 license.

The licenses of the datasets that we used as source do not require consent for using or curating. It is permitted by the corresponding licenses.  

The intended use of DP-Bench is to support research on bridging the gap between natural language intent and structured data representation. It is made publicly and freely available for research purposes at: \url{https://zenodo.org/uploads/20311981}.

\subsection{Ethics Review Board Approval}

The usage of BIRD and ELT-Bench datasets and the subsequent annotations for DP-Bench were reviewed and approved by lawyers of our company.
\end{document}